\documentclass[acmsmall]{acmart}
\usepackage{makecell}
\usepackage{subcaption}
\usepackage{xcolor}
\usepackage[most]{tcolorbox}
\usepackage{multirow, makecell}
\usepackage{tablefootnote}

\AtBeginDocument{%
  \providecommand\BibTeX{{%
    \normalfont B\kern-0.5em{\scshape i\kern-0.25em b}\kern-0.8em\TeX}}}

\setcopyright{acmlicensed}
\acmJournal{PACMHCI}
\acmYear{2024} \acmVolume{8} \acmNumber{CSCW1} \acmArticle{84} \acmMonth{4} \acmPrice{15.00}\acmDOI{10.1145/3637361}

\newcommand{\RNum}[1]{\uppercase\expandafter{\romannumeral #1\relax}}

\begin{document}

\title[Investigating Human-AI Co-creativity in Prewriting with Large Language Models]{``It Felt Like Having a Second Mind'': Investigating Human-AI Co-creativity in Prewriting with Large Language Models}

\author{Qian Wan}
\affiliation{%
  \institution{City University of Hong Kong}
  \city{Hong Kong}
  \country{China}
}
\email{qianwan3-c@my.cityu.edu.hk}

\author{Siying Hu}
\affiliation{%
  \institution{City University of Hong Kong}
  \city{Hong Kong}
  \country{China}
}
\email{siyinghu-c@my.cityu.edu.hk}

\author{Yu Zhang}
\affiliation{%
  \institution{City University of Hong Kong}
  \city{Hong Kong}
  \country{China}
}
\email{yui.zhang@my.cityu.edu.hk}

\author{Piaohong Wang}
\affiliation{%
  \institution{City University of Hong Kong}
  \city{Hong Kong}
  \country{China}
}
\email{piaohwang2-c@my.cityu.edu.hk}

\author{Bo Wen}
\affiliation{%
  \institution{University of Macau}
  \city{Macau}
  \country{China}
}
\email{bowen@um.edu.mo}

\author{Zhicong Lu}
\affiliation{%
  \institution{City University of Hong Kong}
  \city{Hong Kong}
  \country{China}
}
\email{zhicong.lu@cityu.edu.hk}


\renewcommand{\shortauthors}{Qian Wan et al.} 

\begin{abstract}
Prewriting is the process of discovering and developing ideas before writing a first draft, which requires divergent thinking and often implies unstructured strategies such as diagramming, outlining, free-writing, etc. Although large language models (LLMs) have been demonstrated to be useful for a variety of tasks including creative writing, little is known about how users would collaborate with LLMs to support \textit{prewriting}. The preferred collaborative role and initiative of LLMs during such a creative process is also unclear. To investigate human-LLM collaboration patterns and dynamics during prewriting, we conducted a three-session qualitative study with 15 participants in two creative tasks: story writing and slogan writing. The findings indicated that during collaborative prewriting, there appears to be a three-stage iterative \textit{Human-AI Co-creativity} process that includes \textit{Ideation}, \textit{Illumination}, and \textit{Implementation} stages. This collaborative process champions the human in a dominant role, in addition to mixed and shifting levels of initiative that exist between humans and LLMs. This research also reports on collaboration breakdowns that occur during this process, user perceptions of using existing LLMs during \textit{Human-AI Co-creativity}, and discusses design implications to support this co-creativity process.
\end{abstract}



\received{January 2023}
\received[revised]{July 2023}
\received[accepted]{November 2023}

\maketitle

\section{Introduction}
Large language models (LLMs) are machine learning-based language models that are pre-trained on large amounts of data \cite{brown2020language}. Recent models such as GPT-3 have been shown to exhibit high levels of accuracy and performance for tasks \cite{bommasani2021opportunities} such as translation \cite{brown2020language}, question asking \cite{mishra2021cross}, story writing \cite{chung2022talebrush,swanson2021story}, and  programming \cite{weisz2021perfection}. They have also opened up the potential for human-AI collaboration \cite{bommasani2021opportunities}, as such models can quickly adapt to user-specified tasks that are created using only natural language descriptions or prompts. 

Prior research has already shed light on the strategies and challenges that exist when collaborating with LLMs in real-life writing scenarios \cite{singh2022hide,wu2022ai,sun2022investigating}. In particular, the potential to use LLMs for creativity support has become a focus in the academic literature \cite{chung2022talebrush} and among online communities such as Reddit, where there has been an increasing number of users that have collaborated with LLMs to write poets, stories, and even books (e.g., on r/OpenAI \footnote{https://www.reddit.com/r/OpenAI/}, r/GPT3 \footnote{https://www.reddit.com/r/GPT3/}, r/WritingWithAI \footnote{https://www.reddit.com/r/WritingWithAI}, etc.). Compared to other writing tasks, creative writing requires divergent thinking \cite{runco2014creativity,sawyer2011explaining}, which involves the creative generation of multiple answers to a set problem. Existing literature often approaches creative writing in convergent thinking tasks or phases with the goal of reaching a final or at least decent draft \cite{wu2022ai,wu2022promptchainer,swanson2021story,singh2022hide}, leaving earlier divergent thinking phases such as prewriting under-explored.
As a result, this field lacks nuanced insights into key stages of the creativity process \cite{wallas1926art,sternberg1999concept,frich2019mapping} from a human-AI collaboration perspective \cite{wang2020human}.

In prewriting that entails divergent thinking, originality is as equally important as effectiveness \cite{runco2012standard,frich2019mapping}. In such contexts, goals and expectations of human-AI collaboration might differ from convergent thinking phases or tasks. For example, during the pre-writing phase of a science fiction story, writers often begin by generating and organising ideas, asking LLMs to ideate as many novel plots as possible. To this end, the suggestions generated by LLMs might have to be original and diverse in the first place, other than being coherent, plausible, structured, or logically-sound. Furthermore, prewriting is known to be iterative and unstructured \cite{rohman1965pre}. To prewrite with an LLM, one should expect to organise their vague and unstructured thoughts into prompts iteratively, which presents unique collaboration challenges. It remains unknown what challenges might arise when using state-of-the-art LLMs during such process that is iterative and unstructured, and how they in turn affect collaboration patterns for creativity. We formalise these aforementioned gaps as the following research questions:

\begin{itemize}
    \item \textbf{RQ1}: During creative tasks, which collaboration processes, workflows, or strategies are adopted by users while prewriting with LLMs and what is the LLMs' role during this collaboration?
    \item \textbf{RQ2}: What challenges exist when using existing LLMs to support prewriting and how do these challenges affect collaboration?
\end{itemize}

To address these research questions, we conducted a three-session qualitative study with 15 participants. During the study, participants completed common creative tasks, i.e., story writing and slogan writing. The findings uncovered a novel three-stage collaboration process, i.e., \textit{Human-AI Co-creativity}, that exists while prewriting with a state-of-the-art LLM (GPT-3). This process was found to include three distinct stages: \textit{Ideation}, \textit{Illumination}, and \textit{Implementation}. The results also highlighted the dominant role of humans during such collaborations and the mixed and shifting levels of initiative that exist during the \textit{Human-AI Co-creativity} process. Lastly, the findings highlighted common collaboration breakdowns and workarounds that were employed while collaborating with LLMs and led to several design implications that should be considered during the development of future LLMs for creativity support.

Our contributions to the HCI and CSCW community are threefold. First, we provide a nuanced description of the \textit{Human-AI Co-creativity} process, situated in the context of prewriting with LLMs (GPT-3) in two creative tasks. Second, we summarise collaboration breakdowns and workarounds during this process using the existing LLM. Third, we provide design implications for future work to leverage LLMs for creativity support.
\section{Related Work}
The present research was informed and inspired by prior research on creativity, human-AI collaboration, and writing support tools, especially those using LLMs.

\subsection{Creativity Process Models}
Our study is informed by a long line of psychological models of creativity.
In 1929, Wallas presented one of the first models of the creative process that includes four phases: preparation, incubation, illumination, and verification \cite{wallas1926art}. According to Wallas, creative thoughts began with preparation and unconscious processing (i.e., incubation). Creative ideas emerged into conscious awareness during illumination and were ultimately verified, elaborated, and applied during the verification stage. The modern-day creativity research is usually considered to have begun with Guilford's address to the American Psychological Association (APA) in 1950 \cite{sternberg1999concept}, in which he accentuated the need to engage more profoundly in the study of creativity, and to focus attention on scientific approaches to conceptualizing creativity, as Guilford argued that creativity could be studied objectively by examining cognitive processes \cite{runco2014creativity}.

At the intersection of creativity research and HCI, a huge amount of work was dedicated to supporting creativity using computing technologies, notably via the design of Creativity Support Tools (CSTs) \cite{frich2019mapping}.  Early work by Shneiderman proposed a four-phase framework to support the development of user interfaces for creative problem solving, which included phases where one Collects ( learn from previous works), Relates (consult with peers and mentors), Creates (explore, compose, discover and evaluate possible solutions), and Donates (disseminate results) \cite{shneiderman2001supporting}. In 2019, Frich et al. conducted a systematic review of previous CSTs developed by researchers since 1999. Through a thematic analysis of these academic papers, they identified 6 stages of creativity that those CSTs aimed to support, including pre-ideation, ideation, evaluation, implementation, iteration and project management \cite{frich2019mapping}.

To approach today's co-creativity process with an LLM, it is necessary to integrate perspectives from both psychology and creativity support.
The former delineates a human's psychological process without modelling the interaction with computers and intelligent agents, while the latter focuses on available CSTs without referring to the complete psychological process.
To this end, our research conceptualizes the \textit{Human-AI Co-creativity} process. While engaging with Wallas' model, \textit{Human-AI Co-creativity} focuses on collaboration with an intelligent agent beyond human thinking processes alone. Compared to Shneiderman's and Frich's models, our model incorporates the \textit{Illumination} stage from Wallas' model, a stage often overlooked in previous studies of creativity support. We refer our readers to \autoref{three_stage_model} for how \textit{Human-AI Co-creativity} model was developed.

\subsection{Human-AI Collaboration and Co-creativity}
Nowadays, it has become commonplace for humans to collaborate with Artificial Intelligence (AI) for tasks such as decision-making \cite{cai2019human,lee2021human,puranam2021human,jain2022effective}, gaming \cite{zhang2021ideal,ashktorab2020human}, content moderation \cite{lai2022human}, education \cite{ng2020understanding}, and data science \cite{wang2019human}. According to Wang et al., ``collaboration'' is more complex than interaction because it involves ``\textit{mutual goal understanding, preemptive task co-management, and shared progress tracking}'' \cite{wang2020human}. To integrate AI into the complex human workflows, they propose to bring a Computer-Supported Cooperative Work (CSCW) perspective. Echoing this call, recent years have seen a growing number of research that approaches AI as a ``collaborator'' with goals, designated tasks, and abilities to communicate with humans. For example, Wang et al. found that there was a frequent mismatch between AutoAI's goal to produce models of high prediction accuracy and data scientists' goals to understand relationships in data \cite{wang2019human}. To coordinate tasks between human and AI, Mackeprang et al. proposed a method to find an optimal task allocation based on a levels-of-automation (LoA) framework \cite{mackeprang2019discovering}.
 
From early-day research on human interactions with computers \cite{amershi2019guidelines,horvitz1999principles,tecuci2007seven}, to recent advances in human-AI collaboration \cite{rezwana2022designing}, initiative has always been studied as one of the key research questions. The most popular discourse on initiative concerns whether direct manipulation or intelligent interface agents should be employed for user interfaces \cite{farooq2017human,shneiderman1997direct}. The former grants users control and predictability, while the latter requires the delegation of tasks to agents. Research has also proposed a ``mixed-initiative'' principle to support interactions between humans and computers \cite{horvitz1999principles,tecuci2007seven}. To support the design of human-AI co-creative systems, Rezwana and Maher proposed the Co-Creative Framework for Interaction design (COFI), which categorized human-AI collaboration styles into spontaneous or planned based on the timing of initiative \cite{rezwana2022designing} . They found that most existing co-creative systems support ``planned'' instead of ``spontaneous'' timings, suggesting non-improvisational co-creativity.

Along this line of research, recent studies have extensively explored the potential of AI, notably generative models, for creative tasks such as drawing \cite{oh2018lead,yan2022flatmagic,davis2016co,davis2016empirically,lee2011shadowdraw}, contemporary art composition \cite{caramiaux2022explorers}, fashion design \cite{jeon2021fashionq}, musical composition \cite{louie2020novice}, digital mood board creation \cite{koch2020imagesense}, and so on. Several works have specifically studied the co-creation experience from a human-AI collaboration perspective. For instance, Oh et al. investigated co-creation user experiences during drawing tasks, focusing on communication and initiative \cite{oh2018lead}. Their findings in controlled experiments revealed that humans always wanted to lead and preferred ``\textit{just enough instruction}''. In particular, as Natural Language Processing (NLP) technologies have gained traction, research to support creative writing has become increasingly popular \cite{lu2018inkplanner,roemmele2018automated,zhang2022storydrawer,singh2022hide,clark2018creative,gero2019metaphoria,yuan2022wordcraft,gero2022sparks}. Clark et. al explored the potential of human-AI co-creation during creative writing through two machine-in-the-loop prototypes \cite{clark2018creative}, echoing prior work on mixed initiative user interfaces \cite{horvitz1999principles}. They found that users generally expected AIs to deviate from existing results during early stages of story writing and criticised AIs for lacking novelty during slogan writing tasks.

Inspired by this prior research, the present exploration provides nuances on how initiative shifts between users and LLMs and users' preferred collaborative roles during each key stage of the \textit{Human-AI Co-creativity} process in two prewriting tasks.

\subsection{Writing Support and Large Language Models}
Dating back to 1981, Flower and Hayes modelled writing as three on-linear and hierarchical cognitive processes, i.e., planning, translating, and reviewing \cite{flower1981cognitive}. Alternatively, Rohman divided the writing process into three iterative phases that included prewriting, writing, and rewriting \cite{rohman1965pre}. Without explicitly referencing the creativity literature, Rohman \cite{rohman1965pre} connected the prewriting stage to the creativity process by formalising it as ``\textit{that activity of mind which brings forth and develops ideas, plans and designs, not merely the entrance of an idea into one's mind; an active, not a passive enlistment in the `cause' of an idea; conceiving, which includes consecutive logical thinking but much more besides; essentially the imposition of pattern upon experience.}''.

In recent years, with significantly scaled up model sizes, large language models (LLMs) have emerged as a promising tool to support a range of writing tasks \cite{devlin2018bert,brown2020language,wei2022emergent}, including academic writing \cite{wu2022ai,wu2022promptchainer}, story writing \cite{chung2022talebrush}, etc. One of the most successful LLMs, GPT-3, is able to generate high-quality natural language text via natural language description of tasks or prompts \cite{dale2021gpt}. To leverage LLMs for creative writing, Chung et al. proposed a generative story ideation tool using line sketching interactions with an LLM. It supported granular sequence control over the fortune of the protagonist in a GPT-generated story, by translating sketches of users to GPT prompts via a control module \cite{chung2022talebrush}.

Despite their emergent capabilities, LLMs require careful prompt design \cite{wu2022ai}, resulting in a body of literature in NLP dedicated to prompt engineering \cite{lu2021fantastically,liu2021makes,betz2021thinking,reynolds2021prompt} and automatic prompt optimisation \cite{shin2020autoprompt,zhou2022large}. There have also been several efforts in HCI to make LLMs more transparent, explainable, and controllable to support collaboration. For example, Wu et al. introduced the idea of chaining LLM steps together, wherein the output of one step became the input of another \cite{wu2022ai,wu2022promptchainer}. They created a set of LLM primitive operations and proposed a framework for chaining these operations together to produce satisfying results. Sun et al. also investigated the explainability of LLMs for code generation using a question-driven approach \cite{sun2022investigating,liao2020questioning,liao2021question}. They proposed different types of explainable AI features such as AI documentation to support LLM usage in coding scenarios.

Within this body of literature, prewriting was seldom approached as a standalone stage. By exact definition, prewriting is a mental activity before ``writing ideas are ready for words or on the pages'', implying iterative and often loosely-structured workflows \cite{rohman1965pre}. Only a handful of tools, such as \cite{lu2018inkplanner,sadauskas2015mining}, were framed as prewriting tools that specifically support mental activities of idea generation and organisation rather than implementation.  Our work enriches empirical understanding of how users might perceive and leverage an LLM for creative tasks during prewriting, the earliest stage of the writing process. While our study is situated in the context of prewriting, we draw upon creativity models to further dissect the collaborative process. This perspective presents new challenges of modeling uncertainty, as the creativity process entails divergent thinking \cite{runco2014creativity,sawyer2011explaining} and values originality over quality \cite{runco2012standard,frich2019mapping}.

\section{Method}
To investigate human-AI collaboration dynamics during prewriting tasks, we conducted a qualitative study based on Constructivist Grounded Theory \cite{charmaz2006constructing}. Thus, our findings about the \textit{Human-AI Co-creativity} process were co-constructed among researchers, participants' data, and existing theories. The entire study was conducted in Mandarin and was audio and screen recorded.

\subsection{Participants}
Purposive sampling  was used to recruit participants. We recruited students of creativity-related majors at our university (e.g., creative media, art \& design, literature, etc.) and specifically targeted those with little to none expertise in AI research or engineering as we believed that a layman's usage of LLMs would not be biased by any technical details of the model. This process resulted in 15 participants agreeing to participate in our study (\autoref{tab:demographics}; P1-15). All participants were ethnically Chinese and English was their second language. All participants provided consent to participate in the study and agreed to audio and screen recording of the session. Each participant was provided with a coupon equivalent to 50 HKD after the study to thank them for their participation. 

\begin{table*}[!ht]
    \begin{tabular}{cccccc}
    \toprule
    \textbf{ID} & \textbf{Age} & \textbf{Gender} & \textbf{Knowledge of AI} &     
    \textbf{Story Writing Genre}
    & \textbf{Slogan Type}\\
    \midrule
    P1 & 28 & Female & No knowledge & Action fiction & Marketing slogan \\
    P2 & 27 & Male & A little & Science fiction & Marketing slogan \\
    P3 & 28 & Male & No knowledge & Science fiction & Film title \\
    P4 & 20 & Female & A little & Science fiction & Film title\\
    P5 & 20 & Female & No knowledge & Detective fiction & Film title\\
    P6 & 21 & Male & A little & Horror fiction & Social media ad \\
    P7 & 21 & Female & A little & Science fiction & Film title \\
    P8 & 23 & Female & No knowledge & Horror fiction & Marketing slogan \\
    P9 & 23 & Female & A little & Science fiction & Film title \\
    P10 & 22 & Female & A little & Science fiction & Film title \\
    P11 & 25 & Female & No knowledge & Horror fiction & Film title \\
    P12 & 22 & Female & Frequent user & Science fiction & Film title \\
    P13 & 24 & Male & A little & Action fiction & Film title \\
    P14 & 25 & Male & Frequent user & Action fiction & Marketing slogan \\
    P15 & 22 & Female & A little & Science fiction & Marketing slogan \\
    \bottomrule
    \end{tabular}
    \caption{Participant Demographics and the story writing genres and slogan types used during the study.}
    \label{tab:demographics}
\end{table*}

\subsection{Tasks}
The study used two writing tasks: story writing and slogan writing. These tasks were chosen as they were the most common writing tasks mentioned in the literature about human-AI collaboration and creativity support \cite{singh2022hide,clark2018creative,chung2022talebrush}.
Each participant was required to first complete the story writing task about a given scenario such as detective fiction, science fiction, and so forth. In the story writing task, each participant was asked to at least work out and articulate a general storyline. In the slogan writing task, the participant was asked to write a concise but memorable slogan for the story writing fiction they just came up with. The slogan took the form of a film title, a social media advertising post, a marketing slogan, and so on.

\subsection{Study Procedure}
At the beginning of the study, each participant completed a demographic survey about their age, gender, ethnicity, first language, and AI knowledge. They were also asked to list some creative writing tasks they encountered in their daily lives. We then introduced participants to the concept of LLMs and trained them on how to generate and input prompts by walking them through the OpenAI GPT-3 Playground API. In addition to using the examples and tips provided by OpenAI, we also used examples and tips that were collected from online communities (e.g., r/WritingPrompts, r/WritingWithAI, etc.) and the academic literature (e.g., \cite{arora2022ask,dang2022prompt,reynolds2021prompt}; \autoref{apdx_training_matertial}). Each participant was also allowed to explore the interface for 10 minutes.

The study was then comprised of three sessions: a scenario-based ideation session, a think-aloud session, and an interview session. In an earlier pilot study with two HCI researchers, we found that participants became fixated on the affordances of the LLM interface, although in prewriting they could adopt various strategies such as free-writing, listing, mind-mapping, concept mapping, etc. Therefore, we added an ideation session before the usage of LLMs, where each participant would ideate about how they would prewrite with an LLM on a writing interface. 

\subsubsection{Session \RNum{1}: Scenario-based Ideation}
After the training session, each participant was given a pen and a piece of paper. The participant was asked to treat the paper as a writing interface that they could freely write or draw on and prompt the AI to generate ideas whenever necessary. We then assigned the participant a genre from our genre bank (e.g., science fiction, horror fiction, etc.) and asked the participant to ideate on different prewriting strategies that could be used with the LLM for story writing and slogan writing. We used prewriting strategies from previous literature \cite{baroudy2008procedural} as a starting point for the ideation process, which included concept mapping, brainstorming, free-writing, mind-mapping, listing, Q/A (How, When, What, Why), and so forth. We then asked each participant about their initial ideas or thoughts on prewriting with the LLM, and then walked them through existing strategies that could be used as prompts. For each prewriting strategy, we showed participants formal descriptions and example images and asked them how they would collaborate with the LLM if they were to adopt such a strategy. We then asked them to reflect on the strategy, whether they liked it or not, and why. Each participant was also required to demonstrate their prewriting strategies or collaboration workflows by writing or drawing on the paper. 

\subsubsection{Session \RNum{2}: Think-Aloud}
After the ideation session, participants were encouraged to execute a think-aloud implementation of their strategies for the LLM interface. Each participant used the OpenAI GPT-3 Playground interface \footnote{https://beta.openai.com/playground} to try out their prewriting strategies to complete the two writing tasks, in order. We chose this API because it was the state-of-the-art LLM model at the time of our study. Moreover, GPT-3 Playground was flexible enough for prewriting as it offered a variety of options with examples (e.g., `insert mode' to generate in the middle of given inputs, `Q/A' templates for conversation-like interaction, etc.). The freedom of organizing prompts within the interface also overcame concerns about fixation \cite{jansson1991design}. Participants could specify where the AI should write and model parameters such as randomness. 

At the beginning of this session, all GPT-3 parameters used OpenAI's default values (Model: text-davinci-002 model, Temperature: 1, Top P: 1, Frequency penalty: 0, Presence penalty: 0, Best of: 1). Each participant was informed beforehand about the meaning of each parameter and when parameter tuning could be helpful.
Each participant was then required to perform the story-writing task to create a rough, but articulated, idea of a storyline and then perform the slogan-writing task to promote the story. We asked participants to speak about their collaboration workflow, prewriting strategies, how they prompted the LLM, how they perceived the output, and so on. To assist participants in collaborating, we provided prompt guidance using the examples and tips provided during the training session, but only when the collaboration broke down and the participant did not know how to obtain output after several failed attempts of rewriting prompts. During the implementation of the prewriting strategies, we also asked participants to compare their expectations with their experiences and if any mismatches would affect their strategies. After the two tasks, we also asked participants to use the LLM to perform creative writing tasks that they carried out in their daily lives.

\subsubsection{Session \RNum{3}: Interview}
In the third session, we conducted a semi-structured interview in a reflective manner. Participants were first asked to reflect on their usage of the LLM, including the prewriting and prompt strategies they adopted, their general perception of the LLM (e.g., ``What do you think of the creative capability of the LLM?''), overall collaboration experience (e.g., ``What breakdowns did you come across?'', ``What output of the LLM were the most impressive?''), and so on. Based on this reflection, we then asked them to think of potential future designs that could be implemented to support prewriting with the LLM during creative tasks.

\subsection{Data Analysis}\label{three_stage_model}
Our study data was comprised of screen and audio recordings and the strategies and workflows drawn on paper by participants. To analyze this data, the first author (with an AI research background) first performed an initial open coding \cite{corbin2014basics} of the audio and screen recording by playing them simultaneously, and referring to the paper drawings when necessary to understand strategies adopted. During this round of coding, the author focused on what the participant said about the LLM, how he or she collaborated with it, and what he or she said, wrote, or drew about the prewriting strategies. They also recorded timestamps, a description of the data (transcription of audio or description of video content), initial themes that emerged, and kept an analytic memo to track emerging themes.

During a discussion session, the first and second author (with a design background) then analyzed these initial codes through the lens of the existing theories of creativity and writing from the literature. They both concurred that the existing models were insufficient to interpret the data because emerging themes could not fit in one single writing or creativity model. Writing models such as \cite{rohman1965pre,flower1981cognitive} did not further decompose the psychological process of the prewriting stage from a creativity perspective. Creativity theories from Psychology, such as Wallas' four-stage model \cite{wallas1926art}, described human thinking processes without the mention of interactions with computers. Existing models of creativity support also either precluded collaboration with an intelligent agent such as an LLM \cite{shneiderman2001supporting} or lacked key stages in the psychological process, such as illumination \cite{frich2019mapping}. The first and second author then agreed to choose Wallas' \cite{wallas1926art} and Frich's \cite{frich2019mapping} models as the starting point of a new theoretical framework, as these models were thought to be sufficient to cover the psychological processes and state-of-the-art creativity support.

Based on these models, the first and second author performed a second round of coding, where we revisited the analytic memo and assigned the initial themes into different stages of the two models. We then grouped these stages into categories and constructed a new model that had three key stages: \textit{Ideation}, \textit{Illumination}, and \textit{Implementation}. The terminology of \textit{Ideation} and \textit{Implementation} was borrowed from Frich's model, though these two stages were also mentioned by Wallas. In Wallas' model, \textit{Ideation} related to both preparation and incubation, and \textit{Implementation} was included in the verification stage. The second stage, \textit{Illumination}, came from Wallas' model and was rarely touched on by previous CSTs. Finally, we validated our codes by looking at each data description and the recordings and extracted higher-level themes to complement prior theories or construct new theories or insights in the analytic memo. All of the codes and themes were later translated into English for reporting herein.
\begin{figure*}
    \centering
    \includegraphics[width=.9\linewidth]{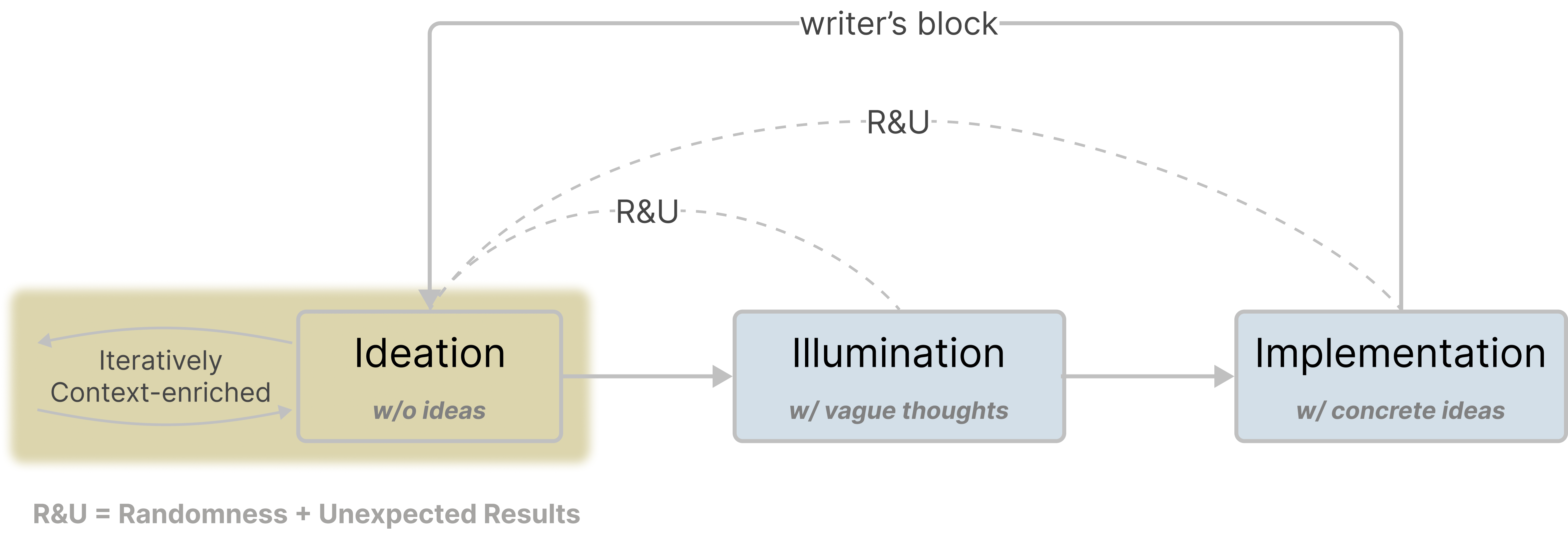}
    
    \caption{The \textit{Human-AI co-creativity} process that exists during prewriting when collaborating with LLMs.}
    \label{workflow}
\end{figure*}

\section{Findings}
Drawing upon previous literature on the creativity \cite{wallas1926art,frich2019mapping} and writing \cite{flower1981cognitive,rohman1965pre} processes, we found that LLMs were most intensively used to support three key stages of the entire creativity process: \textit{Ideation}, \textit{Illumination}, and \textit{Implementation} (\autoref{workflow}). Participants usually used the LLM for \textit{Ideation} when they initially had no ideas or only a vague picture in their minds. If they happened to have any thoughts, most preferred using the LLM as an \textit{Illumination} tool to organize, summarize and reify their existing thoughts, rather than for ideation. Once an idea could be articulated or formalized, participants often experimented with them by writing them down during the \textit{Implementation} stage, either as a title while slogan writing or as a film script or storyline while story writing.

In general, the three stages often occurred in the linear order, where creative thoughts were noticed during \textit{Ideation}, elucidated during \textit{Illumination}, and experimented with during \textit{Implementation}. However, just as previous theories of writing suggest \cite{flower1981cognitive,rohman1965pre}, the process of prewriting is also iterative, notably the usage of LLMs for \textit{Ideation}. We found that participants often used the LLM for \textit{Ideation} whenever they encountered writer's block during the \textit{Implementation} stage. Furthermore, the unexpected results and randomness (sometimes even ``failure'') of the LLM output from any of the three stages was also considered to be a source of inspiration, which implicitly led to \textit{Ideation}.

Herein, we first report on each of the three stages of the creativity process, focusing on participants' prewriting and prompt strategies, and describe the roles of the human and LLMs during collaboration (RQ1). We then report on how participants perceived the LLM during collaboration and common breakdowns and workarounds that occurred during Session \RNum{2} (RQ2).

\subsection{Human-LLM Collaboration During Ideation}
The results demonstrated that the LLM was used in a variety of explicit, iterative, and random ways to help generate ideas during \textit{Ideation}. Examples of such utilization are described next.

\subsubsection{Explicit Ideation}
When participants did not have an initial idea, they would directly ask the LLM for ideas in the hopes that it would provide brief keywords or outlines to spark their inspiration. This was preferred to the generation of longer passages of text that made perfect sense. For example, P5 wanted the LLM to keep randomly generating related keywords or phrases from which she could draw inspiration during slogan writing. P9 instead wanted to ideate with the LLM by listing related keywords or general ideas. She said she could write down some brief keywords for a science fiction story (e.g., outdoor, expenditure) to prompt the LLM to generate related ideas. She would also list some general ideas (or plots) and ask the LLM to follow the pattern of the list and continue to write. During Session \RNum{2}, she wrote ``\textit{1. time travel and save life 2. exploring in the jungle, 3. investigation of evidence}'' as a few-shot prompt, and expected the LLM to continue the list. P8, said she wanted to structure the general settings (e.g., time, location, activities, themes, etc.) of a horror fiction using concept maps and then prompt the LLM to ideate on each concept. She expected the LLM to generate something like ``December'', ``around Christmas'', ``School Life'', etc.
P11 was required to write a ``horror fiction'' in Session \RNum{1}, and she said she would like the LLM to provide a list of horror elements in ancient Chinese cultures at the very beginning to seek inspiration.

\subsubsection{Context-Enriched, Iterative Ideation}
During Session \RNum{2}, we observed that explicit ideation was usually iterative. After the first attempt at ideation, the output of LLM was often perceived to be too general, and somewhat bland, typically because the prompts of the first attempt were also broad and vague.
To improve the results, participants usually added more context to the initial prompt, such as naming a scenario, providing an example, or directly specifying where or how to improve. For instance, after working on a rough plot for a science fiction story, P2 said he would like to improve the output by enriching the ``\textit{emotional arc}''. He added that the hero (protagonist) of the story fell in love with a nurse after waking up in a hospital and asked the LLM to generate what would happen next. The LLM continued, ``\textit{The nurse is killed by the monster}'', which P2 thought was quite amazing. While performing the slogan writing task, we also observed that P10 was constantly changing her prompts by adding constraints. Upon getting the initial output ``\textit{Change is inevitable}'', she thought it was too short, and changed the last sentence of her prompt from ``\textit{Please write a slogan based on the story above.}'' to ``\textit{Please write a slogan in two sentences based on the story above.}''. After submitting the prompt several times, she obtained results such as ``\textit{The women killed the alien: a fresh start for Mars}'', but this time she thought the result was too plain so she tried asking the LLM to improve the results by using rhetoric such as metaphors or exaggeration.

Despite the results being general and bland during Session \RNum{2}, especially during the first few attempts, most participants (e.g., P1, P3-6, P8, P10) mentioned that they would still use LLMs for ideation as long as they could provide something new. P8 explained that she only expected the LLM to \textit{``introduce new concepts that she could not think of''}, and she would \textit{``take care of the rest and process those concepts into intriguing and articulated ideas''}. P4 also stressed in reflection that she believed the LLM could generate more ideas than humans and that it provided ideas that escaped her during collaboration. After prompting the LLM to use a metaphor in the generated slogan, P10 obtained an awkward result that did not look like a slogan at all, i.e., ``\textit{The alien's blood on her hand was a symbol of the couple's new beginning}''. She said it still inspired her about where or how she could draw a metaphor, although the result did not fit the prompt perfectly.

\subsubsection{From Implementation Back to Ideation}
Participants (e.g., P4-6) also often encountered writer's block during the \textit{Implementation} stage while experimenting with their existing ideas, so they would turn to the LLM for another round of ideation. Among our participants, P3 spent most of their time free-writing with the LLM during Session \RNum{2}, mostly in a co-creation-like manner, because he already had an idea of his opening scene. In an explanation of his strategies, he said,
\begin{quote}
    \textit{``I would like to write down something as a scene, and then save it and move to the next scene. I would be using the LLM either to generate a scene or something if I happen to run out of ideas, or as a source inspiration to explore other possibilities.''}
\end{quote}

During his implementation of the story writing task, he first wrote down a simple opening scene for a science fiction film, where the protagonist, John opened up his eyes and managed to lift his upper body. The LLM was then asked to continue the opening so it wrote that a woman showed up in the monitor in front of him and welcomed his return. P3 then further wrote that John asked who she was and where he was and the LLM wrote that the woman had to keep her identity secret and that John was in cryogenic sleep for fifty years before waking up in the space station. P3 said he did not think of such impressive, detailed plot when asking the LLM who the woman was and that he would directly follow the generated ideas to continue writing.

When participants iteratively used the LLM for ideation during \textit{Implementation}, they often had difficulty structuring and designing their prompts because they needed to ask the LLM for ideation based on their already written context. Most participants ended up directly using previous output as the context in the prompt and added an imperative sentence (e.g., ``Please brainstorm something'') or used the `insert mode' of the OpenAI API, which did not always provide satisfying results. While free-writing their opening scene with the LLM, P4 found that the sentence ``\textit{The spaceship is pulled into the black hole and Alan is killed}'' was a little confusing, and she wanted to ask the LLM to brainstorm how Alan was killed. Assuming that adding an imperative sentence within an opening scene would be odd and invalid, she used the `insert mode' several times with the prompt: ``\textit{The spaceship is pulled into the black hole and Alan is killed by [insert]}''. The LLM failed to understand the prompt was intended for brainstorming how Alan was killed and instead generated something similar to ``\textit{The spaceship is pulled into the black hole and Alan is killed by the intense force}''.

\subsubsection{Unexpected Results and Randomness}
We also found that the LLM could be implicitly used for ideation, where ideas were accidentally generated. Unexpected results and the randomness of the LLM were the most common source of inspiration. The unexpected results during Session \RNum{2} were either caused by the ambiguity in the prompts or the limitations of the LLM. In such cases, the LLM returned seemingly valid generations that amazed the participant, but derailed them from their original intentions. For instance, during P8's slogan writing, the LLM generated a title that was thought to be out-of-context, i.e., ``\textit{the demonic dough}''. However, P8 said it is intriguing because it provided a new concept, ``dough'' that was ``demonic'', which she did not think of while story writing. She could have considered adding more suspense or horror elements related to this concept, e.g., writing a scene about a bakery lesson where dough was being made and something supernatural happened. While free-writing, P2 expected the LLM to specify the upbringing of the hero (protagonist) of his story so he prompted the LLM by asking for the ``background'' of the hero based on a given storyline. This ambiguous prompt led the LLM to generate the background of the story rather than the hero alone, such as where monsters came from, how the monsters destroyed the world, and how humans fought back led by the hero. P2 said the results were impressive and he was open to continuing  ideation or writing based on such generation.

\begin{figure*}
\begin{subfigure}{.32\textwidth}
    \centering\captionsetup{width=\linewidth}
    \includegraphics[width=.95\linewidth]{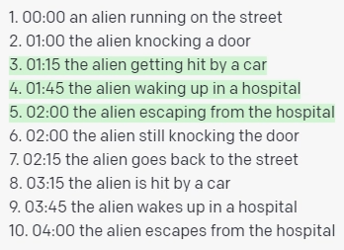}
    \subcaption{The randomness of the LLM's output, which was perceived as inspiring by P10.}
    \label{P10}
\end{subfigure}
\begin{subfigure}{.6\textwidth}
    \centering\captionsetup{width=.8\linewidth}
    \includegraphics[width=.95\linewidth]{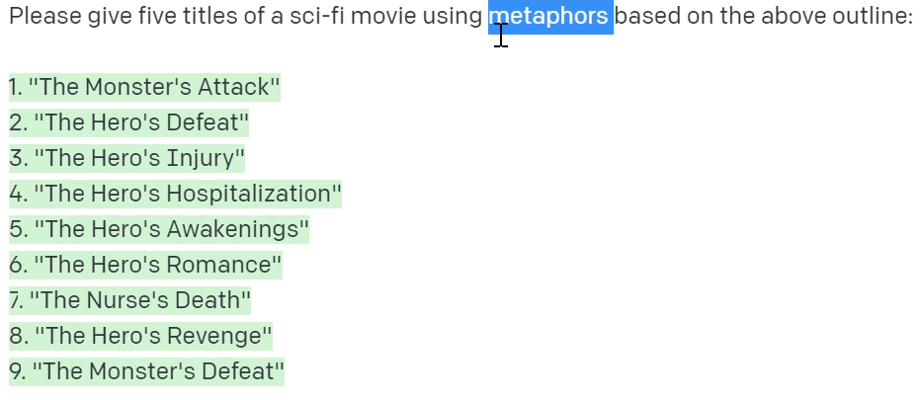}
    \subcaption{The collaboration broke down between P2 and the LLM when the LLM generated nine titles that did not contain a metaphor.}
    \label{P2}
\end{subfigure}
\begin{subfigure}{\textwidth}
    \centering\captionsetup{width=.75\linewidth}
    \includegraphics[width=.975\linewidth]{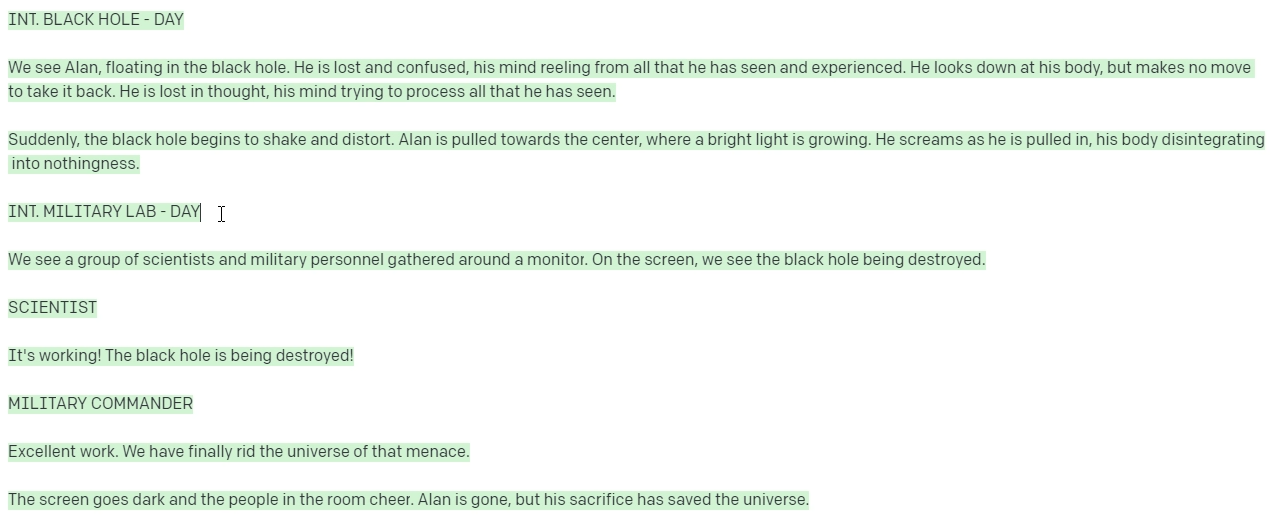}
    \subcaption{P4's LLM-generated science fiction film script, which was perceived to be vivid in details, but have a corny ending.}
    \label{P4}
\end{subfigure}
\end{figure*}

Implicit ideation was also triggered by the randomness of the LLM, where participants drew inspiration from LLM output that was believed to be a complete failure or breakdown. While listing storylines in the form of timelines using the LLM's ``insert mode'', P10 expected the LLM to continue writing what happened at a given time. She then came across incorrect output where the LLM only repeated what was already written by P10 at another time, which she said was nonsensical ( \autoref{P10}). However, P10 said that such output was quite inspiring as it reminded her about using a time loop in her storyline where an alien was trapped and managed to break out.

\subsection{Human-LLM Collaboration During Illumination}
Illumination is a stage where creative ideas burst into conscious awareness and become articulated or formalised. Collaboration with LLMs was often found to involve summarization, organization, and reification of vague, rough, and disjointed thoughts. We now report how participants leveraged the LLM for illumination and how useful it was as a tool for illumination.

\subsubsection{Leveraging LLMs for Illumination}
Many participants (e.g., P4, P6, P8-9, P12-15) mentioned that if they already had something in mind, they preferred using the LLM to illuminate their existing thoughts rather than for ideation. Such ``thoughts'' were mostly in incubation, which participants found difficult to consciously articulate. While writing a science fiction story, P9 said she already had some rough ideas in mind, but she was only able to name some vague keywords such as ``outdoor'', ``expenditure'', ``starry night'', and so on. During Session \RNum{1}, she said she preferred using concept maps and asking the LLM to organise keywords and generate other details such as characters, locations, plots, and so on. For example, the LLM could merge the two concepts such as ``outdoor'' and ``starry night'' in the concept map and generate possible plots in new concept boxes that could be added to the existing concept map. She also noted that she might sometimes free-write some fleeting thoughts on the prewriting interface and she expected that the LLM would keep track of her thoughts and present a summary or a suggested storyline. Similarly, P7 mentioned that she would like to list some thought fragments and expected the LLM to organise them into something promising.

When using the LLM to summarise and organise their existing thoughts, participants expected the LLM to help elucidate their ideas with nuances via the reification of vague and fleeting concepts. For example, P8 said she thought of a ``\textit{terrible night}'' after \textit{Ideation} with the LLM, and wanted the LLM to keep track of such thoughts and provide some clarification such as what terrible things happened that night. P10 said that while listing their vague and unstructured thoughts with the LLM, she expected the LLM would help enrich existing ideas in the list with more details.
P12 initially outlined elements of a story such as characters, locations, themes, events, etc., and asked the LLM to provide more details such as relationships between characters, possible settings for scenes, how an event unfolds, etc.

While slogan writing during Session \RNum{2}, P4 created a title for her science fiction film where an astronaut was pulled into a black hole and accidentally discovered the truth of the universe and ``\textit{experienced everything that has ever happened and will ever happen}'' before his death. She thought of keywords such as ``lights'' to represent the experience of the astronaut and expected the LLM to use them in a concise title, preferably via rhetoric (e.g., use the word ``lights'' as a contrast against the black hole). Therefore, she directly used a previously written storyline as context to ask the LLM to name the story within 6 words using the list of keywords.

\subsubsection{The Capabilities of LLMs as an Illumination Tool}
The capabilities of the LLM were often noted as being astonishing, where vague prompts with fragmented concepts could be interpreted by the LLM into concrete and fascinating ideas. Some participants, such as P2, commented that ``\textit{the AI seemed to know what I was about to say}'', while they could not articulate their ideas themselves. P4 used the LLM to draft a content plan for a real-life mobile app design task. She only had a vague idea of the immersive mobile application she wanted to design, which monitored users' heart rate and potentially featured summer night scenes and fireflies. She said the app could be a sports or health app, or a game, but it was far from even a prototype. She then prompted the LLM in a quite messy way,
\begin{quote}
   \textit{``Please write an overview of my application which includes a philosophical description of fireflies and life, summer night scenes, and heart rate frequency of the user."}
\end{quote}
The LLM then articulated a concrete idea that had her in awe,
\begin{quote}
    \textit{``The application is a digital recreation of a summer night scene, with fireflies flitting about and the user's heart rate affecting the frequency of their flashes. The philosophy behind the fireflies is that they represent life -- with its ups and downs, its surprises and miracles, its beauty and fragility. Just as each firefly's light is unique, so too is each person's life. And just as the fireflies' light fades as summer comes to an end, so too will our time on this earth come to an end. But the light of our lives can still shine brightly and bring joy to others, even after we're gone."}
\end{quote}
P4 said the LLM seemed to understand what was in her mind, articulated the function of the application, and abstracted philosophical meaning based on fragmented concepts via a very concrete and detailed description. She said that she came across the idea when she saw a meadow during the summer. She suddenly imagined fireflies and wanted to incorporate such a scene into the application, i.e., ``\textit{it was a crazy idea that didn't seem to make sense, but the AI makes it work.}''

\subsection{Human-LLM Collaboration During Implementation}
Implementation occurred after participants already had a concrete idea that was small and wasn't accompanied by a bigger picture. They then tentatively experimented with their ideas by writing them down and polishing them with the LLM. During Session \RNum{2}, there were generally two types of collaboration to implement an idea, i.e., participants directly asked for LLM generations based on given specifications of the idea or they prompted the LLM to fill in details based on what they wrote themselves. 

The first approach was usually used at the beginning of the \textit{Implementation} stage. Participants liked to clarify their structured ideas as the context to tentatively prompt the LLM for a first draft. For example, during Session \RNum{2}, P8 specified characters, locations, time, themes, and key events about her story and directly asked the LLM to write an overview of the complete storyline. P5 instead asked the LLM to generate an opening scene of a detective film script based on her plot.

The second approach was generally adopted by participants to polish their writing. Based on a draft, the LLM was asked to bridge a logical gap in a plot, provide a nuanced portrait of a scene, or rephrase a paragraph. For example, after adding a love story between the hero and the nurse, P2 wanted more details about how the emotional arc affected the main storyline of the fight against a monster. He used the `insert mode' to prompt the LLM to generate content within his storyline and obtained the suggestion ``\textit{The nurse is killed by the monster}''. While free-writing with the LLM, P6 felt the following scene lacked nuance and could be further enriched:
\begin{quote}
    \textit{They went on a bus bound to a village. But when they jumped on the bus happily, they suddenly found that was the last bus today. Worse, all passengers got off three bus stops before reaching the terminus.}
\end{quote}
He therefore added an ``[insert]'' between ``\textit{...the last bus today}'' and ``\textit{Worse, ...}'' to prompt the LLM to provide more details to explain why all passengers got off early.

When implementing their ideas, most participants (e.g., P2-6) noted that they would like to take care of the general picture, either a storyline or the rhetoric of a slogan. The LLM was expected to enrich the details or fill the gaps. For instance, P3 explained that \textit{``I would like to scaffold the writing by providing a structure and a general storyline ... I would use it mostly for details or to overcome writer's block''}. He also added that it was mainly just his own preference.
However, this behavior was found to be quite common among our participants and was often caused by perceptions of the LLM after extended usage during Session \RNum{2}. The LLM's ability to generate nuanced details was appreciated by many, but its ability for high-level logic reasoning or sense-making was often questioned. During Session \RNum{2}, P6 tried prompting the LLM to enrich or develop his main storyline many times, but most of the LLM generations were viewed as failures to understand his general context. The LLM initially generated ``\textit{Worse still, the bus was out of service}'', which was contradictory to the context. 

During their interviews, P2, P3, P4 and P5  spoke highly of the LLM because of its capability of generating nuanced details, while they all acknowledged the LLM seemed to have problems understanding their general ideas or developing a storyline in a valid way. In one scenario, the LLM generated a film script based on P4's storyline (\autoref{P4}):

\begin{quote}

    INT. BLACK HOLE - DAY
    
    \textit{We see Alan, floating in the black hole. He is lost and confused, his mind reeling from all that he has seen and experienced. He looks down at his body, but makes no move to take it back. He is lost in thought, his mind trying to process all that he has seen...}
    
    INT. MILITARY LAB - DAY
    
    ...

    \textit{The screen goes dark and the people in the room cheer. Alan is gone, but his sacrifice has saved the universe.}
    
\end{quote}

P4 said the output was vivid in detail but that the ending scene was unexpected and corny. She thought the film should be themed around the discovery of the truth of the universe but the LLM wrote a super-hero style film script.

\subsection{The Collaborative Roles of Humans and LLMs}
In this subsection we summarise the roles of human and the LLM during the whole prewriting process (\autoref{CollaborativeRoles}).

{\renewcommand{\arraystretch}{1.1}
\begin{table*}[htb]
    \centering
    \begin{tabular}{l|l|l|l}
        \toprule
          & \textbf{Scenario} & \textbf{Goal} & \textbf{Initiative} \\
         \midrule
         \textbf{Ideation} & without ideas & generate ideas & LLM-lead \\
         \textbf{Illumination} & with vague thoughts & elucidate thoughts & Human-led \\
         \textbf{Implementation} & with concrete ideas & experiment with ideas by writing & Human-led \\
         \bottomrule
    \end{tabular}
    \caption{Collaboration patterns of \textit{Human-AI Co-creativity} found during the study.}
    \label{CollaborativeRoles}
\end{table*}
}

\subsubsection{The Shift of Initiatives}
Most of the time, participants wanted to take initiative, especially during the \textit{Implementation} stage, unless they ran out of ideas. Only when they had no ideas or they happened to encounter writer's block, would they let the LLM take the initiative to generate ideas from scratch. For example, during Session \RNum{2}, P3 explained that, \textit{``I liked to collaborate with it while I still had control. I could delete generations I didn't like, and ask it [the LLM] to re-generate or simply write on my own. It felt like having a second mind in parallel that processed all the context and provided new ideas when requested.''} P2 also believed that the initiative of the human while collaborating with the LLM was a major advantage compared to human-human collaboration. He said that, \textit{``Human collaborators might have their own unique thoughts and failed to get what I'm thinking or writing but the LLM could do exactly what I asked it to do based on my thoughts.''}

Nevertheless, while letting the LLM take initiative, participants were quite open-minded as long as the LLM could generate something new. During iterative ideation, many participants (e.g., P3-4, P8-9) mentioned they would like the LLM to ``\textit{defend}'' their generations by providing specifications, even if they were vague and confusing in the first place. For instance, during story writing, P8 mentioned she did not mind following the LLM's suggested storylines for ideation if she encountered writer's block, even if the suggested ideas diverged from her expectations. P4 used the LLM to brainstorm an ending for her story and she thought the results were a little vague and seemingly out of line with her context. She said she would like to ask the LLM to explain how it could align its results with the given plot.

\subsubsection{Verification of Results}
As the definition of creativity implies both originality and effectiveness \cite{runco2012standard,frich2019mapping}, the verification of the effectiveness of ideas is also important for creativity support \cite{frich2019mapping}. Throughout the entire creativity process, participants always verified the results themselves and integrated them with their existing thoughts, which later led to an iterative ideation or iteration across prewriting stages.
While asking the LLM to explain its generations, P4 liked to evaluate the results in the following way:
\begin{quote}
  \textit{There are three points to consider. First, does it align with what I expect? Second, what's the difference between my expectation and the LLM generation, and is it acceptable? Third, if the generation is novel and intriguing, how should I integrate it with my existing ideas.}
\end{quote}
In only two cases did we find that participants would use the LLM to verify their ideas. P8 said she would use the LLM as Grammarly to proofread her writing. P2 said he would like the LLM to perform a plagiarism check of his writing against its training data.

We also found that participant agency during collaboration alleviated copyright concerns because participants either took care of the big picture or verified and integrated the generations with their own thoughts if the LLM took the lead. P3 said he would not consider copyright a problem because he was leading the collaboration most of the time. P7 also mentioned copyright was not an issue because she would blend the generations with her own writing rather than use them directly.

\subsection{Breakdowns and Workarounds}
The collaboration breakdowns during prewriting fell roughly into two categories, i.e., dynamically-adjusted context and uncertainty in communication via prompts. These breakdowns were mainly caused by the nature of LLMs rather than affordances of the OpenAI interface.Herein, we report on these breakdowns, how participants worked around them, and design fearures that were suggested to overcome these breakdowns.

\subsubsection{Dynamically-Adjusted Contexts}
Prewriting with LLMs was almost never observed to only require one single prompt. It was usually iterative, which required participants to dynamically adjust the context of their prompts. In this situation,  participants often had no idea what they should do. For example, while moving to slogan writing, many participants (P2-4, P6-7) found that their previous results while story writing were messy and cluttered, which could not be directly used as context. Some participants (P3, P6) had to delete all the results and ask the LLM to brainstorm slogans, sometimes based on a completely rephrased storyline. P4 iteratively asked the LLM to brainstorm endings for her story. She reflected that she had difficulty asking the LLM to enrich each ending on the list with more details and inquired about how each ending connected to the opening. Sometimes she even thought of completely dropping the previous context and starting all over again to brainstorm something for her story.

To support the dynamic adjustment of context, some participants (e.g., P2, P7, P10) mentioned that they wanted to see some examples, suggestions, templates, or tutorials as a reference. For instance, P10 said she preferred a template so that she could fill in the blank to specify her context. P7 said she wanted to have multiple panels in the interface for context adjustments so she could develop the main storyline in a main panel and ask the LLM questions or request details in another panel. In this case, the context in each panel would not be conflated.

\subsubsection{Uncertain Communication via Prompts}
The communication between participants and the LLM often broke down due to the uncertainty of prompt-based communication, which is a known challenge when interacting with generative AI models such as GPT-3 \cite{bommasani2021opportunities}. Many participants reported they did not know how to tell the LLM to do certain things. The most common breakdowns were failure (i.e., nonsensical results) and fixation (i.e., the LLM continued to generate unsatisfactory results regardless of how a prompt was adjusted). In an extreme case, P2 found that minor grammatical mistakes in his prompts could lead to random LLM generations.

Moreover, the LLM could also easily misunderstood or completely ignore the requirements of the participant, even if it generated something seemingly valid. P6 recalled that,
\begin{quote}
  \textit{I was quite curious sometimes what kind of a prompt was needed to get what I want... Some of my prompts were completely ineffective that the LLM failed to understand. I felt it wouldn't work either if I rephrased my prompts and changed some keywords. I'd like to know what prompts are preferred by the LLM.}
\end{quote}

P2 summarised such a situation as ``\textit{context-sensitivity}'', where the LLM could sometimes be extremely sensitive to some words even if they were insignificant in the prompt, and other times completely ignore some words as if they did not understand them. He reflected that the LLM seemed to misunderstand what a slogan was, but returned decent results when he changed ``slogan'' to ``title''. He then once prompted the LLM to \textit{``Please give five titles of a sci-fi movie using metaphors based on the above outline"} the LLM understood what the ``title'' was, but generated nine titles without using metaphors as if it did not understand what ``metaphor'' meant (\autoref{P2}).

During their interview, P2 said that,
\begin{quote}
   \textit{While collaborating with a human, you can communicate with ambiguity and he or she could still understand you but the LLM seemed to comprehend some words by exact definition. You have to be very specific when communicating with the LLM, which could be difficult.}
\end{quote}

He therefore suggested that he would like to see all the potential training data that contributed to the understanding of some keywords or concepts.

\subsection{Perceptions of Existing LLMs}
Participants had various perceptions about existing LLMs during prewriting and varied thoughts on how such perceptions might influence collaboration.

\subsubsection{Strengths}
In general, participants felt that the LLM was good at introducing new concepts during \textit{Ideation}, elucidating vague thoughts with nuance for \textit{Illumination}, and enriching results of \textit{Implementation} with details. Therefore, during the \textit{Illumination} and \textit{Implementation} stage, participants often expected nuanced and detailed results from the LLM while they preferred taking care of the big picture. During the \textit{Ideation} stage, however, results were often said to lack nuance, but could still introduce new concepts which benefited creativity output.

\subsubsection{Mediocrity}
Throughout different stages of usage, almost all participants encountered results that were mediocre. They described them as ``\textit{too general}'' (P1, P14), ``\textit{featureless}'' (P4, P9, P15), ``\textit{too plain}'' (P2, P6, P13-14), ``\textit{corny}'' (P3-4, P15), and so forth. Such mediocre results were partially due to prompts being too vague and broad or too specific. The former lacked proper context to deliver decent results, while the latter often caused the LLM to keep summarizing the existing context. For example, P14 once prompted the LLM to write a slogan based on a detailed storyline, but the results were said to be only a summary of what was given without anything inspiring. P15 instead was initially requesting a background of a science fiction with very vague descriptions, and found results were too general. She guessed ``\textit{perhaps it (GPT-3) couldn't know where to start, or how to return something promising with so little context given}''

Such mediocrity sometimes compromised credibility to the degree that participants would become suspicious of the creativity of the LLM. P15 said that, ``\textit{I feel that it (GPT-3) can only learn from previous data and generate similar patterns, but is unable to conceive anything truly novel, like many masterpieces.}''. Many participants (e.g., P2-3, P9-10) noted they would like to look at the LLM's training data to understand its creative potential.

\subsubsection{High-Level Sense-Making}
Participants often questioned the comprehension skills of the LLM as it often failed to comprehend abstract concepts or high-level themes such as storyline. In these cases, the communication often broke down due to misunderstandings. Therefore, some of participants (e.g., P3, P6) mentioned that they would take care of the big picture themselves. P6 stressed he would not use the LLM in the future for anything ther than details or ideation because it could not understand what he was writing.

\subsubsection{Training Data}
Many participants mentioned that they wanted to see the LLMs training data because they did not trust it. P3 and P4 said the ``corny'' plots generated by the LLM made them feel like it was trained on novels from the last century. P3 added that he would not even consider using it, if the LLM was only trained on old-fashioned fictions.

P1, P9, and P11 said that the plot generated by the LLM was Western-style, which might imply bias in its training data. When asking the LLM to write an action film in ancient times, the LLM generated an opening scene of a king's army on an expedition, while P1 expected Asian styles such as Kung Fu, swordsman, etc. P11 was asking the LLM to search for elements of horror in ancient Chinese culture, but she found LLM was misinterpreting some ancient Chinese stories such as ``Butterfly Lovers'' as ``horror fictions''.
 
`P2 also mentioned that he was suspicious about if the LLM was directly copying its training data, and therefore wanted to look at all data related to its current generations. P5 initially doubted the LLM's ability to writing dialogue in a film script based on the presumption that the LLM might not find dialogue in its dataset.
\section{Discussion}
Our findings lead to a three-stage collaboration process of prewriting in two creative tasks and uncovered current practices that are employed while using an existing state-of-the-art LLM. We first formalize the \textit{Human-AI Co-creativity} process to outline the role of LLMs in generating new concepts and providing nuance and detail how initiative shifts during collaboration. We then situate our model in existing literature of human-AI collaboration and creativity support, and explore its distinctive nature to leverage uncertainty for creativity. We also discuss the design implications of supporting co-creativity from the perspectives of prompt strategies, writing, and explainability. We conclude by documenting the limitations of the \textit{Human-AI Co-creativity} model.

\subsection{Human-AI Co-Creativity Process: Collaborative Roles and Initiatives}
We aim to approach creativity from the perspective of human-AI collaboration. Our findings led to a three-stage collaboration process while prewriting a creative story and slogan that included \textit{Ideation}, \textit{Illumination}, and \textit{Implementation}. We term this process \textit{Human-AI Co-creativity}. It complements previous investigations into the role of LLMs in the writing process \cite{gero2022sparks,yuan2022wordcraft,singh2022hide} by highlighting the collaborative roles and shifting initiatives that exist between humans and LLMs in creativity during prewriting. Our study suggests that during the human-LLM collaboration, LLMs were delegated the tasks of generating new concepts and providing nuance by elucidating vague thoughts and polishing writing with details. In general, when beginning with no or several concrete ideas, LLMs assisted humans as a source of inspiration and reified and enriched their existing thoughts.

Our study also highlighted the dominant role of the human throughout the process of co-creativity, which echoes prior work on human-AI co-creation \cite{oh2018lead}. Taking such a dominant role was two-fold. On the one hand, humans take the generated output from LLMs with a grain of salt. They like to verify the results themselves and integrate output from LLMs with their own thoughts. On the other hand, humans like to take charge of the general picture, such as a storyline or the rhetoric of a slogan. LLMs are mostly used to provide nuanced details, either for elucidation during \textit{Illumination} or while experimenting with their ideas during \textit{Implementation}. However, our work distinguishes itself from prior work \cite{oh2018lead,clark2018creative,chung2022talebrush} as participants did not mind following the LLM for \textit{Ideation}, and even iteratively fine-tuned its results, which benefits divergent thinking \cite{runco2014creativity,sawyer2011explaining}. Furthermore, the \textit{Human-AI Co-creativity} process is an iterative process, which aligns with existing writing models \cite{flower1981cognitive,rohman1965pre}. Apart from iteration led purely by humans, the unexpected results and randomness of LLMs could accidentally revert the three stages back towards \textit{Ideation}.

With such collaboration patterns, LLMs are put in a new position with respect to initiative \cite{farooq2017human,shneiderman1997direct}. During the \textit{Ideation} stage of the co-creativity process, LLMs were granted initiative since the originality of generations was prioritised over quality. For the other two stages, LLMs took more of an assistive role, while the shift of initiatives might accidentally occur based on users' evaluations of collaboration outcomes. In this situation, participants also perceived their outcome, either a slogan or a storyline, as original, due to their agency during collaboration and the relatively lightweight usage of LLMs compared to other LLM writing tasks \cite{wu2022ai,wu2022promptchainer,chung2022talebrush,singh2022hide}. We believe the conceptualisation of \textit{Human-AI Co-creativity} could tentatively help piece together the puzzle of ethical interrogations of AI co-creativity or AI-generated art \cite{dee2018examining,gangadharbatla2022role,lima2021social}.

\subsection{Theoretical Implications: Collaboration and Creativity}
The proposed \textit{Human-AI Co-creativity} process model combines perspectives from human-AI collaboration, creativity theories, and creativity support in two creative writing scenarios. In comparison to previous studies of LLM-powered writing support, such as Singh et al. \cite{singh2022hide}, our model concerns collaboration patterns and dynamics, including goals, initiatives, communication, workflows, etc. This approach aligns with the call for a CSCW perspective \cite{wang2020human} in human-AI collaboration, as the LLM was conceived as an intelligent collaborator throughout our study, both reflected in participants' account and by how we interpreted our data.

Furthermore, the \textit{Human-AI Co-creativity} process enhances previous understanding of human-AI co-creative experience \cite{gero2022sparks,oh2018lead,clark2018creative}, by incorporating Wallas' creativity theories in the two prewriting tasks. In particular, the \textit{Illumination} stage has received limited attention in previous literature of creativity support. To our best knowledge, the closest accounts in HCI to this stage are convergent phases in CSTs (e.g., \cite{jeon2021fashionq}), or iteration of design prototypes \cite{dow2010parallel,dow2009efficacy}. They are both related to refining existing ideas (or prototypes), while the \textit{Illumination} stage starts from unconsciousness or inarticulate thoughts in incubation, and emphasizes the burst of creative ideas into conscious awareness. Previously AI models might not be able to directly translate vague, rough, or disjointed concepts into concrete or promising outcomes, but our findings suggest the feasibility of using LLMs to support \textit{Illumination}.

This new finding can hopefully inspire future design of CSTs to facilitate emergence of creative ideas in incubation. Notably, some theories of creativity~\cite{campbell1960blind,simonton2022blind} posit that a process of \textit{blind variations} is essential for reaching true creativity, during which creators become ignorant of the effectiveness of any creative ideas. In this sense, supporting \textit{Illumination} is crucial in that it can significantly accelerate the \textit{blind variation} process: not by generating alternatives, but ``guiding'' the creator with AI throughout the thinking process. We expect empirical studies to enhance understanding of this stage while collaborating with generative AI models. Specifically, other creative scenarios, such as drawing or music composition, might have different forms of an ``idea'' compared to writing. It requires a new study to figure out how other generative AI models can help ``illuminate'' these forms of ideas.

\subsection{Uncertainty as Creativity in Human-AI Co-Creativity}
While other writing or human-AI collaboration tasks mainly expect certain and precise output \cite{wu2022ai,wu2022promptchainer,sun2022investigating}, one of the most distinctive features of \textit{Human-AI Co-creativity} in the two prewriting tasks was treating uncertainty as a source of creativity. While prewriting with LLMs, participants did not expect a high-quality or once-for-all result that perfectly fits the given prompt. Instead, the LLM was usually used to avoid writer's block or fixation \cite{jansson1991design}, where imperfect ideas, unexpected results, and the randomness of the LLM all served as a source of inspiration provided that they introduced new concepts or remind users of other possibilities. A change of initiative can also occur without irritating users due to the nature of divergent thinking \cite{runco2014creativity,sawyer2011explaining}, which echoes prior work on generative AI models for art and resonates with the call for an inclusive view of AI \cite{caramiaux2022explorers}.

Such uncertainty does compromise the quality and efficiency of human-AI communication and sometimes leads to collaboration breakdowns. This duality of uncertainty in the creativity process advances the exploration of the imperfection of generative AI models \cite{weisz2021perfection,sun2022investigating}, and also opens up possibilities for future explainability features to model uncertainty for creative tasks. We expect future investigations into what kind of, or what level of, certainty could be used for creativity and how we could explain or communicate such uncertainty to facilitate communication and sense-making for creativity support.

Furthermore, the previous literature on LLMs for human-AI collaboration was largely dedicated to controlling the uncertainty of an AI model \cite{wu2022ai,wu2022promptchainer}. Based on our findings, research attention should go beyond reducing randomness or delivering satisfying yet definite generation when it comes to creativity support. During prewriting with a LLM, uncertainty could be leveraged to facilitate divergent thinking by sometimes disrupting communication, although future effort is needed to understand the balance between the two. As the most common participant frustration was the ``mediocre'' generations that seemed reasonable but lacked originality, such needs will pose a challenge for algorithmically improving LLMs or addressing related collaboration breakdown.


\begin{table*}[htb]
    \centering
    \resizebox{\textwidth}{!}{\begin{tabular}{llp{4in}}
        \toprule
        \textbf{Category} & \textbf{Applied to} & \textbf{Definition} \\
        \midrule
        \multirow{2}{*}{Examples \& Tutorials} & \multirow{2}{*}{Inputs} & Example prompts, templates, or related tips needed to address collaboration breakdown \\
        & & \\
        \multirow{2}{*}{Data} & \multirow{2}{*}{Models} & Transparency of task-related training data and those that contribute to the understanding of certain concepts \\
        & & \\
        Context Sensitivity & Outputs & Which part of the prompt most significantly led to the output \\
        & & \\
        \multirow{2}{*}{Capability} & \multirow{2}{*}{Models} & Whether the LLM is capable of understanding or generating something because it has seen related data \\
        \bottomrule
    \end{tabular}}
    \caption{Categories of AI documentation for LLMs for creativity support}
    \label{XAI}
\end{table*}

\subsection{Design Implications}
Several design implications arose from our research relating to prompt strategies for prewriting, writing support tools, and potential explainability features.

\subsubsection{Prompt Strategies for Prewriting}
As prewriting is iterative, prewriting with LLMs cannot be completed via a single prompt without proper context. The context to prompt a LLM also often needs to be dynamically adapted based on user requirements. Thus, there is a design opportunity to track context during collaboration and semi-automate the process of adjusting context by providing suggestions or guidance. For example, systems could prompt users with real-time summaries of context (i.e., what was written; e.g., \cite{dang2022beyond}) and example questions (i.e., how to ask the LLM to do something), whenever the collaboration seems to breakdown. Systems could then adapt off-the-shelf prompt strategies (e.g., \cite{arora2022ask}) to wrap these requirements into a valid prompt. Like one participant (P7) suggested, systems should maintain multiple writing panels in parallel to maintain context. In this case, any iterations, either intentionally or unexpectedly, would not influence the previous context or writing results.

\subsubsection{LLM-Augmented Writing Support Tools}
Our study also shed light on possible writing support tool designs that integrate LLMs. Building upon previous prewriting tools such as \cite{lu2018inkplanner}, design efforts should be made to scaffold the co-creativity process using LLMs. For \textit{Ideation}, for instance, prewriting tools could support translating diagrams or lists into prompts to request more LLM generated concepts. To support \textit{Illumination}, systems could provide features to organize writing in an interface, or elucidate keywords in a diagram. We also expect future tools leveraging LLMs to support creative thoughts in incubation while users step away from a problem. For \textit{Implementation}, systems can also introduce human-LLM co-creation to experiment ideas like previous creative writing systems \cite{chung2022talebrush,yuan2022wordcraft,clark2018creative}.

In addition, the initiative of the human should be properly supported. During a typical ideation scenario, LLMs should be allowed to take the initiative to overcome writer's block or design fixation \cite{jansson1991design}. Prompt strategies should be designed to support iteratively requesting that LLMs polish their generations. In other scenarios, however, humans should be granted full agency while LLMs should take a more assistive role, addressing details without disrupting the creative thoughts of humans. To seamlessly integrate the two initiative modes into one writing tool, efforts should be made to identify the uncertainty of LLMs (or alternatively, the perception of humans) that caused a change of initiative.

\subsubsection{Explainability for Co-Creativity}
Our findings implied that, while working with a LLM, users would rarely care about the technical details of the model. The most requested explainability features in our study used to be related to input and output, such as transparency of training data, guidance of prompt strategies, examples and templates as a guidance for input, etc. We advocate for new dimensions of AI documentation \cite{arnold2019factsheets,hind2020experiences,knowles2021sanction} related to our participants' concerns to support human-LLM co-creativity. We summarize categories of AI documentation identified in our study in \autoref{XAI} for non-expert users to support creativity. The table is adapted from Sun et al.'s templates for AI documentation of generative AI models for code \cite{sun2022investigating}.

\subsection{Limitations and Future Work}
It is important to note that the presented \textit{Human-AI Co-creativity} model was based on two prewriting tasks (i.e., story writing and slogan writing) and involved participants with design or writing related majors and little AI expertise. Our research also used the GPT-3 model. Thus, although some of our findings such as shifting initiatives echo and complement previous research in similar scenarios \cite{clark2018creative,gero2022sparks,yuan2022wordcraft,singh2022hide}, additional efforts are needed to investigate the generality of the process to other creative scenarios (e.g., drawing), writing stages or tasks, expert or novice writers, and LLMs aside from GPT-3. An expert fiction writer, for example, might not follow the \textit{Human-AI Co-creativity} workflow but instead start from a known storyline. With the release of ChatGPT or even GPT-4, some of our findings regarding perceptions or collaboration breakdowns might not apply to such chat bot-style interactions and new collaboration opportunities or challenges might arise. Furthermore, our three-session study protocol might have impacted the collaboration workflow investigated, though it aimed to simulate a general prewriting scenario. Without being introduced to established strategies or asked to both ideate and implement strategies, participants might have employed different workflows in this one-shot study.

It is also worth noting that our participants were ethnically Chinese and English was their second language. This demographic setup allowed us to provide a non-eurocentric documentation of \textit{Human-AI Co-creativity} and opens up future research opportunities into the potential bias of LLMs across ethnic and language groups. Yet, it also means that several findings regarding prompt design or the evaluation of LLM output might need to be taken with a grain of salt for other language groups or native English speakers. For instance, though our participants reported the LLM had cultural misunderstandings, it would require a future study to investigate its overall comprehension of ancient Chinese cultures beyond just a few cases, and many other studies to understand its cultural bias across different language groups.

Nevertheless, this research paves the way for future efforts to revisit creative writing or creativity support from a \textit{Human-AI Co-creativity} perspective to examine applicability of the model to other workflows or scenarios. We also highlight the need for new models or theories relating to more diverse tasks and scenarios and the participation of users with more diverse backgrounds to enrich our understanding of the \textit{Human-AI Co-creativity} process.
\section{Conclusion}
This research presented a three-stage, iterative process of \textit{Human-AI Co-creativity} that was based on the findings from two creative prewriting tasks. It implies that the human plays a dominant role while discovering and developing ideas, but also that there is a shifting initiative between humans and LLMs while collaboratively prewriting. We also reported common breakdowns and user perceptions of LLMs that existed. This research invites future investigation into \textit{Human-AI Co-creativity} and benefit researchers endeavoring to leverage human-AI collaboration for creativity support, such as the design of LLM-augmented writing tools.

\section{Acknowledgement}
We are thankful to all reviewers for their constructive suggestions that helped make this paper much stronger.

\bibliographystyle{ACM-Reference-Format}
\bibliography{references}

\clearpage
\appendix
\onecolumn
\section{Appendix}

\subsection{Training Material: Prompt Strategies}
\label{apdx_training_matertial}

The training material used in this study was comprised of a list of prompt strategies for GPT-3 collected from online resources and the academic literature \cite{arora2022ask,dang2022prompt,reynolds2021prompt}.
Specifically, we divided these strategies into three categories: \emph{Samples}, \emph{Tips}, and \emph{Guidelines} (\autoref{tab:promptStrategies}).

\setcounter{table}{0}
\renewcommand{\thetable}{A\arabic{table}}
\begin{table*}[htb]
       \resizebox{\textwidth}{!}{\begin{tabular}{lp{2.75in}l}
        \toprule
        Category & Description & Content/Example \\
        \midrule
        \multirow{4}{*}{Samples} & 
        Sample results where proper constraints or context were given in the prompt to deliver structural output, often multiple times towards the final writing  \cite{gwern2022gpt, dang2022prompt}
            & \multirowcell{4}[0pt][l]{a 4chan greentext story \footnotemark[1] \\ poetry writing  \footnotemark[2] \\ a short play \footnotemark[3] \\ a Ramen shop story \footnotemark[4]} \\
        \hline
        \multirow{3}{*}{Tips} & 
        Particular words or phrases to be added to the prompt that might significantly improve the quality of the output \cite{kojima2022large}
            & \multirowcell{3}[0pt][l]{"let's think step by step" \footnotemark[5] \\ "tl;dr" for summary \footnotemark[6] \\ stop words/sequences \footnotemark[7]} \\
        \hline
        \multirow{3}{*}{Guidelines} & 
        \multirow{3}{*}{The prompt design tutorials for the GPT-3 model}
            & \multirowcell{3}[0pt][l]{giving instructions \footnotemark[8] \\ describing what you have in mind \footnotemark[9] \\ adding context to the prompts \footnotemark[10]} \\
            & & \\
            & & \\
        \bottomrule
    \end{tabular}}
    \caption{Prompt Strategies for Training}
    \label{tab:promptStrategies}
\end{table*}

\footnotetext[1]{https://absolutewrite.com/forums/index.php?threads/gpt-3-text-generator-excellent-for-story-ideas.352606/}
\footnotetext[2]{https://www.gwern.net/GPT-3}
\footnotetext[3]{https://twitter.com/GanWeaving/status/1585358381191086080}
\footnotetext[4]{https://ricardodejong.com/how-i-used-a-i-to-create-my-first-book-and-generate-a-podcast-from-it/}
\footnotetext[5]{https://medium.com/merzazine/prompt-design-gpt-3-step-by-step-b5b2a7a3ea85}
\footnotetext[6]{https://towardsdatascience.com/gpt-3-parameters-and-prompt-design-1a595dc5b405}
\footnotetext[7]{https://help.openai.com/en/articles/5072263-how-do-i-use-stop-sequences}
\footnotetext[8]{https://towardsdatascience.com/gpt-3-parameters-and-prompt-design-1a595dc5b405}
\footnotetext[9]{https://www.reddit.com/r/WritingWithAI/comments/ybskwc/how\_to\_start\_a\_prompt\_with\_instruction\_or/}
\footnotetext[10]{https://towardsdatascience.com/gpt-3-parameters-and-prompt-design-1a595dc5b405}

Based on these strategies, the training material contained examples such as:

\begin{itemize}
    \item ``Write an one act play with Jesus and Carl Jung''
    \item ``Write a funny and philosophical 4chan greentext story''
    \item ``What is the meaning of life? Let’s think step by step''
    \item ``Brainstorm solutions to increase sales at your store''
    \item ``Come up with ideas for a new product that is environmentally friendly''
    \item ``My company produces reusable water bottles that can be refilled from the tap''
\end{itemize}

This training materials was also used to provide guidance when collaboration broke down and participants had no ideas.

\end{document}